\documentclass[conference,letterpaper]{IEEEtran}
\usepackage{graphicx,color}
\usepackage{cite}
\usepackage{setspace} 
\usepackage{amsmath}
\usepackage{amsmath,amsthm,amssymb}
\usepackage{multirow}
\usepackage{hhline}
\usepackage{epsfig}
\usepackage{subfigure}
\usepackage{epstopdf}
\usepackage{verbatim}
\usepackage{algorithm}
\usepackage{algorithmicx}
\usepackage{algpseudocode}
\usepackage{etoolbox}
\usepackage{cases}
\usepackage{csquotes}
\usepackage{mathtools,stmaryrd}
\SetSymbolFont{stmry}{bold}{U}{stmry}{m}{n}
\usepackage{bbm}
\usepackage{lettrine}
\usepackage{mathrsfs}

\newcommand{\suchthat}{\;\ifnum\currentgrouptype=16 \middle\fi|\;}

\makeatletter
\newcommand*{\indep}{%
  \mathbin{%
    \mathpalette{\@indep}{}%
  }%
}
\newcommand*{\nindep}{%
  \mathbin{%                   % The final symbol is a binary math operator
    \mathpalette{\@indep}{\not}% \mathpalette helps for the adaptation
  }%
}
\newcommand*{\@indep}[2]{%
  \sbox0{$#1\perp\m@th$}%        box 0 contains \perp symbol
  \sbox2{$#1=$}%                 box 2 for the height of =
  \sbox4{$#1\vcenter{}$}%        box 4 for the height of the math axis
  \rlap{\copy0}%                 first \perp
  \dimen@=\dimexpr\ht2-\ht4-.2pt\relax
  \kern\dimen@
  {#2}%
  \kern\dimen@
  \copy0 %                       second \perp
} 
\makeatother

\makeatletter
\newcommand*{\algrule}[1][\algorithmicindent]{%
  \makebox[#1][l]{%
    \hspace*{.2em}% <------------- This is where the rule starts from
    \vrule height .75\baselineskip depth .25\baselineskip
  }
}

\newcount\ALG@printindent@tempcnta
\def\ALG@printindent{%
    \ifnum \theALG@nested>0% is there anything to print
    \ifx\ALG@text\ALG@x@notext% is this an end group without any text?
    \else
    \unskip
    \ALG@printindent@tempcnta=1
    \loop
    \algrule[\csname ALG@ind@\the\ALG@printindent@tempcnta\endcsname]%
    \advance \ALG@printindent@tempcnta 1
    \ifnum \ALG@printindent@tempcnta<\numexpr\theALG@nested+1\relax
    \repeat
    \fi
    \fi
}
\patchcmd{\ALG@doentity}{\noindent\hskip\ALG@tlm}{\ALG@printindent}{}{\errmessage{failed to patch}}
\patchcmd{\ALG@doentity}{\item[]\nointerlineskip}{}{}{} % no spurious vertical space
\makeatother

\allowdisplaybreaks

\begin{document}

\title{On the Deployment of RIS-mounted UAV Networks}

\author{\IEEEauthorblockN{Anupam Mondal$^{*}$, Priyadarshi Mukherjee$^{\dagger}$, and Sasthi C. Ghosh$^{*}$}

\IEEEauthorblockA{$^{*}$Advanced Computing
 and Microelectronics Unit, Indian Statistical Institute, Kolkata 700108, India\\
$^{\dagger}$Department of Electrical Engineering and Computer Science, Indian Institute of Science Education and Research Bhopal\\
Emails: cs2304@isical.ac.in, priyadarshi@ieee.org, sasthi@isical.ac.in}}

\maketitle
\begin{abstract}
Reconfigurable intelligent surfaces (RIS) enable smart wireless environments by dynamically controlling signal propagation to enhance communication and localization. Unmanned aerial vehicles (UAVs) can act as flying base stations and thus, improve system performance by avoiding signal blockages. In this paper, we propose a gradient ascent and coordinate search based method to determine the optimal location for a system that consists of a UAV and a RIS, where the UAV serves cellular users (CUs) and the RIS serves device-to-device (D2D) pairs. In particular, by optimizing the net throughput for both the D2D pairs and the CUs, the suggested method establishes the ideal location for the RIS-mounted UAV. We consider both line of sight (LoS) and non-LoS paths for the RIS and UAV to calculate the throughput while accounting for blockages in the system. The numerical results show that the proposed method performs better than the existing approaches in terms of both the net throughput and the user fairness. 
\end{abstract}


\section{Introduction}
\noindent In recent years, wireless data usage has grown rapidly and is expected to increase more than five times between 2023 and 2028 \cite{ericsson}. In this context, reconfigurable intelligent surfaces (RISs) have emerged as a promising solution to tackle this challenge by ‘controlling’ the wireless propagation environment \cite{basar2019wireless}. An RIS is essentially an array of passive elements placed on a flat metasurface. Unlike traditional methods that adapt to changes in the wireless channel, an RIS actively controls the wireless environment to improve signal transmission. This is achieved by adjusting the properties of its passive elements. Moreover, since an RIS only reflects the incoming signals in a desired direction, it does not require any radio frequency (RF) chains \cite{tewes2022irs}. 

On the other hand, high-frequency signals such as millimeter waves (mmWaves) \cite{ghafoor2020millimeter} are widely used for high-speed data transfer in the device-to-device (D2D) communication. Although mmWaves-based communication is well suited for short-distance D2D communication, it comes with its own set of challenges such as losing signal strength quickly when passing through obstacles and experiencing high signal loss over long distances. These limitations make it difficult to always maintain a stable and strong connection between a user pair, especially in situations where the direct line of sight (LoS) is weak or not good enough to support mmWave communication. In such scenarios, RIS-assisted D2D communication can help obtain the indirect LoS when the direct LoS is blocked \cite{sau2024drams}. 

Using an RIS, the signal can be intelligently reflected and redirected to the desired destination, even when obstacles block the direct path. This helps ensure a more reliable and efficient wireless connection. Furthermore, to make communication more reliable, RIS is placed in a strategic location where they have a clear LoS with the users who want to communicate with each other. This is crucial, since there is a significant difference in between the LoS and non-LoS (NLoS) path loss (PL) in mmWave communication. NLoS communication occurs when buildings or other obstacles block the direct signal path, leading to higher signal loss and weaker connections. Moreover, this strategic placement helps to improve signal quality and overall network performance \cite{zeng2020reconfigurable,kang2023double,sau2025priority}. In this setup, the signal from one user is reflected by an RIS placed within its communication range before reaching the intended receiver. This improves network connectivity, especially in areas where the direct LoS link is weak or blocked by obstacles.

In addition, for situations such as natural disasters and/or temporary high-traffic areas, unmanned aerial vehicle (UAV)-aided communication can be a very useful alternative, where the traditional existing infrastructure may be unavailable or overloaded. UAVs are widely used in such cases because they offer high flexibility, quick and easy deployment, making them an effective solution for emergency and temporary communication needs \cite{wu2021comprehensive}. As a result, UAVs, like RIS, can aid mmWave communication systems to avoid signal blockages and significantly reduce the associated PL \cite{gapeyenko2021line}. Since UAVs can move freely and adjust their positions, they can find the best locations to maintain a LoS link, ensuring a more stable and reliable connection. In this context, a potential RIS-aided UAV network can deal with the blockage problem more efficiently \cite{uavris1}.  Here the UAV can act as a flying base station to cellular users (CUs) and the RIS on top of it can help establish LoS links between the mmWave-based D2D user pairs. Because the RIS-assisted UAV network serves both types of users, this increases the system throughput. But in that case, the placement of the RIS-mounted UAV is a very challenging issue. Unlike the thoroughly investigated research direction of optimal UAV \cite{curef} and RIS \cite{d2dref} positioning, here we need to consider the net performance of both the CUs and the D2D pairs. It is important to note that the optimal UAV position corresponding to the CUs may or may not be the same as the optimal RIS position corresponding to the D2D pairs, and vice-versa. 
However, since the RIS is mounted on the UAV, they have to be in the same location. Thus, in this work, we investigate the deployment aspect of RIS-assisted UAV networks, where our objective is to enhance the combined system throughput of both the CUs and the D2D pairs. More specifically, our contributions can be summarized as follows:

\begin{itemize}
\item First, we find the optimal position $r_D$ of the RIS for maximizing the throughput of the D2D pairs based on gradient ascent, which may or may not be the best position with respect to the CUs.
\item Next, we find the optimal position $r_C$ of the UAV to maximize the throughput of the CUs, which again, may or may not be the best position for the D2D pairs.
\item Finally, we find the combined optimal location of the RIS-mounted UAV based on coordinate search, by using $r_D$ and $r_C$, which aims to maximize the net throughput of both the D2D pairs and the CUs.
\end{itemize}
We perform extensive simulations to demonstrate the
benefits of the proposed method over the existing benchmark schemes, which individually optimizes the performance of the D2D pairs or the CUs. More precisely, our approach outperforms existing approaches in terms of both net throughput and user fairness. Rest of the paper is organized as follows. Section \ref{system} discusses the system model and Section \ref{prop_strat} presents our proposed method for finding the ideal location of the RIS-mounted UAV. The simulation results are presented in Section \ref{simulation}. Finally, concluding remarks are given in Section \ref{conclusion}.

\section{System Model}\label{system}
\noindent Here we discuss both the considered network topology as well as the mathematical construction of the problem.

\begin{figure}[!h]
    \centering
    \includegraphics[width=\linewidth]{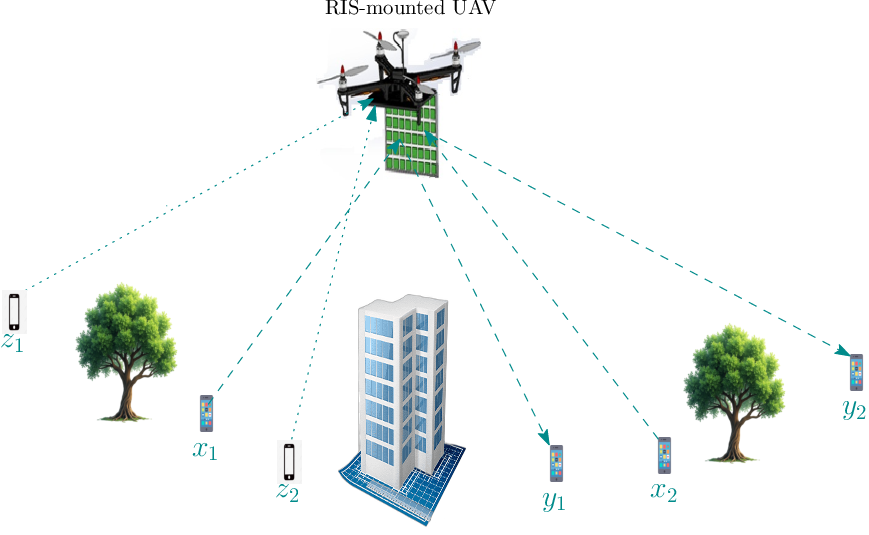}
    \vspace{-2mm}
    \caption{RIS-mounted UAV communication model.}
    \vspace{-4mm}
    \label{smodel}
\end{figure}

\subsection{Network Topology}
\noindent We consider a mmWave wireless system with $N$ active CUs, $M$ D2D pairs, and a RIS-mounted UAV that operates in the orthogonal frequency division multiple access (OFDMA) mode \cite{ofdma}. The RIS is made up of $R$ reflecting patches and the UAV flies at a fixed altitude $H$ above the ground level \cite{height}. Also, we assume that the RIS is able to obtain complete channel state information (CSI) and that the wireless channels corresponding to the consecutive reflecting elements are independent and identically distributed. We denote $\mathscr{C} = \{1, 2, \ldots, N\}$ and $\mathscr{D} = \{1, 2, \ldots, M\}$ as the index sets for active CUs and D2D pairs, respectively. In addition, we assume $\{ (x_1, y_1), (x_2, y_2), \ldots , (x_M, y_M) \}$ is the set of $M$ D2D pairs communicating via the RIS and $\{ z_1, z_2, \ldots , z_N \}$ is the set of $N$ CUs, which communicate with the UAV as a base station. So, we have $K = 2M + N$ devices that communicate simultaneously. Fig. \ref{smodel} demonstrates the scenario for two CUs $z_1,z_2$ communicating with the UAV and two D2D pairs $(x_1, y_1)$, $(x_2, y_2)$ communicating via the RIS.

\subsection{Throughput Calculation for D2D Pairs}

\noindent For $x_m$ $\forall$ $m=1,\ldots,M$, the received signal at $y_m$ is
\vspace{-2mm}
\begin{equation}  \label{d2d}
    s_{y_m}=\sqrt{P_{y_m}} \left[\sum\limits_{\zeta=1}^R |h|_{m,\zeta}e^{-j\psi_{m,\zeta}}a_{\zeta}|g|_{m,\zeta}e^{-j\phi_{m,\zeta}} \right]s_m+ n_o,
\end{equation}
where $P_{y_m}$ is the path loss dependent received power, $s_m$ is the transmitted signal, and $n_o \sim \mathcal{CN}(0,N_0)$ denotes the additive white Gaussian noise (AWGN). Here, $a_{\zeta}=\omega_{\zeta} e^{j\theta_{\zeta}}$ $\forall$ $\zeta=1,\ldots,R$ is the reflection coefficient of the $\zeta$-th RIS reflecting element and $\omega_{\zeta}(\theta_{\zeta})$ denotes the amplitude (phase) adjustment factor of the same. Without any loss of generality, we consider $\omega_{\zeta}=1$ $\forall$ $\zeta$ \cite{sau2024drams}. Moreover, $|h|_{m,\zeta} (\psi_{m,\zeta})$ and $|g|_{m,\zeta} (\phi_{m,\zeta})$ denote the amplitude (phase) of the channel coefficient of the $x_m-R$ link and the $R-y_m$ link, respectively. Furthermore, $P_{y_m}$ depends on the PL model considered \cite{li2022geometric} as follows.
\vspace{-2mm}
\begin{equation}  \label{pymdef}
    P_{y_m} [{\rm dBm}]= P_{x_m}+G_{x_m}+G_{y_m}-PL_{x_m,RIS}-PL_{RIS,y_m}.
\end{equation}
Here, $P_{x_m}$ is the transmit power of $x_m$, $G_{x_m}(G_{y_m})$ is the gain of the transmit (receive) antenna, and
\begin{align}  \label{ploss}    PL_{x_m,RIS}&=\alpha_{x_m,RIS}+10\beta_{x_m,RIS}\log_{10}(d(x_m,r)) \:\: {\rm and} \nonumber \\
    PL_{RIS,y_m}&=\alpha_{RIS,y_m}+10\beta_{RIS,y_m}\log_{10}(d(r,y_m)),
\end{align}
where $ \alpha $ and $ \beta $ are parameters of the LoS and NLoS models respectively, $d(x_m,r)$ is the Euclidean distance between $x_m$ and RIS, and $d(r,y_m)$ is the Euclidean distance between RIS and $y_m$. Lastly, $|h|_{m,\zeta}$ and $|g|_{m,\zeta}$ follow the Rician or Rayleigh distribution, depending on whether they correspond to the LoS or NLoS channel, respectively\cite{sau2025priority}. As we assume the RIS to have the complete CSI, the resulting optimal signal at $y_m$ is
\vspace{-2mm}
\begin{equation}
    s_{y_m}=\sqrt{P_{y_m}} \left[\sum\limits_{\zeta=1}^R |h|_{m,\zeta}|g|_{m,\zeta} \right]s+ n_o
\end{equation}
and the associated total throughput for all the D2D pairs is
\vspace{-2mm}
\begin{equation}  \label{dd2d}
    D_{\rm D2D}=\sum\limits_{m=1}^M \log_2 \left(1+\frac{P_{y_m} \Big| \sum\limits_{\zeta=1}^R |h|_{m,\zeta}|g|_{m,\zeta} \Big|^2}{N_0}\right).
\end{equation}

\subsection{Throughput Calculation for CUs}
\noindent For $z_n$ $\forall$ $n=1,\ldots,N$, the received signal at the UAV is
\vspace{-2mm}
\begin{equation}
    s_{z_n}=\sqrt{P_{n}}f_n s_n + n_o,
\end{equation}
where $P_{n}$, analogous to $P_{y_m}$ in \eqref{d2d}, is the distance dependent received power, i.e., it is a function of $d(z_n,r)$ as follows.
\vspace{-2mm}
\begin{equation}  \label{pndef}
    P_{n} [{\rm dBm}]= P_{z_n}+G_{z_n}+G_{UAV}-PL_{z_n,UAV}.
\end{equation}
Here, $P_{z_n}$ is the transmit power of $z_n$, $G_{z_n}(G_{UAV})$ is the gain of the transmit (UAV) antenna, and
\begin{equation}  \label{plosscu}
    PL_{z_n,UAV}=\alpha+10\beta\log_{10}(d(z_n,r)).
\end{equation}
Also, $f_n$ is the corresponding complex channel gain, and $s_n$ is the transmitted signal. Accordingly, the associated total throughput for all the CUs is given by
\vspace{-2mm}
\begin{equation}    \label{dcu}
    D_{\rm CU}= \sum\limits_{n=1}^N \log_2 \left(1+\frac{P_{n} |f_n|^2}{N_0}\right).
\end{equation}
Similar to $|h|_{m,\zeta}$ and $|g|_{m,\zeta}$ of the D2D scenario, $|f_n|$ is Rician/Rayleigh distributed depending on the wireless channel. In this manuscript, we investigate the RIS-mounted UAV deployment, where we intend to maximize the net throughput $D_{\rm net}$, which is defined from \eqref{dd2d} and \eqref{dcu} as
\begin{equation}    \label{dnet}
    D_{\rm net}=D_{\rm D2D}+D_{\rm CU}.
\end{equation}
\vspace{-7mm}
\section{Proposed Strategy}\label{prop_strat}
\noindent In this section, we introduce a step-by-step strategy to maximize \eqref{dnet}, which is  the net throughput of the CUs and the D2D users. Specifically, we intend to individually maximize $D_{\rm D2D}$ and $D_{\rm CU}$ as two different sub-problems, by finding the best possible solution for each. Subsequently, by taking the solutions from both and combining them, we solve the original optimization problem. The motivation for this approach lies in the fact that, since the RIS is mounted on the UAV, we cannot have different locations for the RIS and the UAV in order to serve the D2D pairs and the CUs, respectively. 
%Moreover, maximizing $D_{\rm net}$ does not necessarily imply maximizing both $D_{\rm D2D}$ and $D_{\rm CU}$.
%\vspace{-3mm}
\subsection{Throughput Maximization for D2D Pairs}

\noindent Now, we obtain insights on the optimal placement of the RIS, which results in the maximization of $D_{\rm D2D}$. Here, $D_{\rm D2D}(r)$ from \eqref{dd2d}, can be expressed as a function of the position of the RIS, i.e., $r = (x_r, y_r, H)$ as
\vspace{-2mm}
\begin{equation}  \label{abcd}
    \!\!D_{\rm D2D}(r)\!=\!\!\sum\limits_{m=1}^M \log_2 \!\!\left(\!1+\frac{P_{y_m}(r) \Big| \sum\limits_{\zeta=1}^R |h|_{m,\zeta}|g|_{m,\zeta} \Big|^2}{N_0}\!\right),
\end{equation}
where from \eqref{pymdef}, we have
\vspace{-2mm}
\begin{equation*}
P_{y_m}(r) [{\rm dBm}]= P_{x_m}+G_d-PL_{x_m,RIS}-PL_{RIS,y_m},
\end{equation*}
$G_d=G_{x_m}+G_{y_m}$, and $PL_{x_m,RIS},PL_{RIS,y_m}$ as stated in \eqref{ploss}. Note that, while $P_{x_m}$ and $G_d$ are constants, $PL_{x_m,RIS}$, $PL_{RIS,y_m}$ depend on $r$ and hence, $P_{y_m}$ is a function of $r$. Thereafter,  by appropriate linearization (i.e, conversion from dBm to Watt) of $P_{y_m}(r)$ and denoting $\Big| \sum\limits_{\zeta=1}^R |h|_{m,\zeta}|g|_{m,\zeta} \Big|^2 = \kappa_m$, \eqref{abcd} can be rewritten as 
\begin{align}  \label{dmax} 
   \!\!\!\!\!\! & D_{\rm D2D}(r) = \sum\limits_{m=1}^M \log_2 \left(1+\frac{P_{y_m}(r) \kappa_m}{N_0}\right) \nonumber\\
   \!\!\!\!\!\! &= \!\!\sum\limits_{m=1}^M \log_2 \left(1+{\eta_m d(x_m,r)^{-\beta_{x_m,RIS}}d(y_m,r)^{-\beta_{RIS,y_m}}}\right),
\end{align}
where $\eta_m = \frac{\kappa_m}{N_0} 10^{\frac{P_{x_m}+G_d-\alpha_{x_m,RIS}-\alpha_{RIS,y_m}-30}{10}} $. Accordingly, by using \eqref{dmax}, we state the following optimization problem.
\vspace{-2mm}
\begin{equation}    
{\rm P1:} \quad \underset{r}{\max} \quad  D_{\rm D2D}(r) \label{max} 
\end{equation}
subject to  $x_{r_{\min}} \leq x_r  \leq x_{r_{\max}}$ and $y_{r_{\min}} \leq y_r \leq y_{r_{\max}}$. 
% \begin{align*}
%      \quad x_{r_{\min}} &\leq x_r  \leq x_{r_{\max}},\\
%      \quad y_{r_{\min}} &\leq y_r \leq y_{r_{\max}}.
% \end{align*}
To solve $\rm P1$, we first calculate $ D_{\rm D2D}'(r)=\frac{\partial D_{\rm D2D}(r)}{\partial r}$ to obtain \eqref{dderi}. %Next, we set it to zero for obtaining $r$ from this equation.
\begin{figure*}
\begin{align}    \label{dderi}    
    D_{D2D}'(r)  &= \sum\limits_{m=1}^M \frac{-\eta_m A_m}{(\ln{2})\left(1+{\eta_m d(x_m,r)^{-\beta_{x_m,RIS}}d(y_m,r)^{-\beta_{RIS,y_m}}}\right)}, \:\: \text{where}
\end{align}
$A_m=\beta_{x_m,RIS}d(x_m, r)^{-\beta_{x_m,RIS}-1}d'(x_m, r) d(y_m, r)^{-\beta_{RIS,y_m}} + \beta_{RIS,y_m}d(x_m, r)^{-\beta_{x_m,RIS}}d(y_m, r)^{-\beta_{RIS,y_m}-1} d'(y_m, r).$
\vspace{0.8mm}
\hrule
\vspace{-2mm}
\end{figure*}
%Note that, when computing the gradient, we must select the LoS and NLoS parameters appropriately, for both the links $x_m-RIS$ and $RIS-y_m$. 
However, if we set $D_{D2D}'(r)=0$ and try to solve for $r$, we do not get a closed-form solution. Therefore, we employ a numerical technique to find the best possible value of $r$. 

Specifically, as $\log_2 (\cdot)$ is a continuous and concave function, it has a global maximum point. Hence, we propose a gradient ascent \cite{gradasc} based algorithm, which will converge to this maximum point. Accordingly,  Algorithm \ref{algo1} solves $\rm P1$, where we avoid $r = x_m$ and $ y_m $ for all $ m = 1,2,\dots,M $. We simply cannot take the gradient at  $r = x_m $ and $ y_m $ for all $ m = 1,2,\dots,M $ because the objective function is not defined at these points. To handle this, we define ${\rm sign}(x)=+1$ for $x>0$ and $-1$, elsewhere. Also, a small tolerance $ \epsilon $ is used, i.e., if $d(x_m,r)$ or $d(r,y_m)$ is less than $ \epsilon $, we adjust $ r $ by moving it away from $ x_m $ and $ y_m $ $\forall$ $ m $ by a displacement factor $\delta$. Furthermore, the bounds of $ r $ are carefully maintained by continuously adjusting its value after each update, making sure that it remains within the specified limits. In \eqref{max}, we determine the limits of $r$ as $r_{\min} = (x_{r_{\min}}, y_{r_{\min}}, H)$, $r_{\max} = (x_{r_{\max}}, y_{r_{\max}}, H)$, and we denote the initial value of $r$ as $r_{0} = (x_{r_{0}}, y_{r_{0}}, H)$. Here $r_{\min}$ and $r_{\max}$ are obtained so that after these points the PL is greater than the PL at $r_0$ \cite{li2022geometric}. Finally, the algorithm stops if 1) The derivative of $D_{D2D}(r)$ evaluated at any iteration is less than $\epsilon$, or 2) The distance between two consecutive locations is less than $\epsilon$, or 3) The maximum number of iterations is reached. The worst case time complexity of this algorithm is $O(MN_{\text{max}})$.
\begin{algorithm} [!h]
\caption{Algorithm to Solve $\rm P1$}  \label{algo1}
\begin{algorithmic}[1]
    \State \textbf{Initialize} with initial guess $r_0$ such that $r_{\min} < r_0 < r_{\max}$ and $r_0 \neq x_m, y_m$ $\forall$ $m$, learning rate $\alpha$, tolerance $\epsilon$, and maximum iterations $N_{\text{max}}$.
    \State Set iteration count $k = 0$.
    \Repeat
        \State Compute $D_{\rm D2D}'(r)$ by using \eqref{dderi}.
        \State $r_{k+1} = r_k + \alpha D_{\rm D2D}'(r_k)$
        \ForAll{$m$}
            \If{$|r_{k+1} - x_m| < \epsilon$}
                \State $r_{k+1} = r_{k+1} + \operatorname{sign}(r_{k+1} - x_m) \cdot \delta$
            \EndIf
            \If{$|r_{k+1} - y_m| < \epsilon$}
                \State $r_{k+1} = r_{k+1} + \operatorname{sign}(r_{k+1} - y_m) \cdot \delta$
            \EndIf
        \EndFor
        \State Ensure bounds: $r_{k+1} = \min(\max(r_{k+1}, r_{\min}), r_{\max})$
        % \begin{equation}
        % r_{k+1} = \min(\max(r_{k+1}, r_{\min}), r_{\max})
        % \end{equation}
    \Until{$|D_{\rm D2D}'(r_{k+1})| < \epsilon$ or $|r_{k+1} - r_k| < \epsilon$ or $k \geq N_{\text{max}}$}
    \State \Return $r^* = r_{k+1}$ and $D_{\rm D2D}(r^*)$
\end{algorithmic}
\end{algorithm}

\subsection{Throughput Maximization for CUs}
\noindent Now we define an optimization problem specifically for the purpose of maximizing $D_{\rm CU}$. Here, $D_{\rm CU}(r)$ from \eqref{dcu}, is expressed as
\vspace{-2mm}
\begin{equation}
    D_{\rm CU}(r)= \sum\limits_{n=1}^N \log_2 \left(1+\frac{P_{n}(r) |f_n|^2}{N_0}\right),
\end{equation}
where from \eqref{pndef}, we obtain
%\vspace{-2mm}
\begin{equation}
    P_{n} [{\rm dBm}]= P_{z_n}+G_c-PL_{z_n,UAV},
\end{equation}
$G_c=G_{z_n}+G_{UAV}$, and $PL_{z_n,UAV}$ as stated in \eqref{plosscu}. By performing similar algebraic manipulations as in Section III-A, we obtain
\vspace{-2mm}
\begin{equation}
    D_{\rm CU}(r)=\sum\limits_{m=1}^M \log_2 \left(1+{\lambda_n d(z_n, r)^{-\beta_{{z_n,UAV}}} }\right),
\end{equation}
where $\lambda_n=\frac{|f_n|^2}{N_0}10^{\frac{P_{z_n}+G_c-\alpha_{z_n,UAV}-30}{10}}$. Similar to \eqref{max}, we frame the following optimization problem.
\vspace{-2mm}
\begin{equation}     \label{maxc}
{\rm P2:} \quad \underset{r}{\max} \quad  D_{\rm CU}(r)
\end{equation}
subject to  $x_{r_{\min}} \leq x_r  \leq x_{r_{\max}}$ and $y_{r_{\min}} \leq y_r \leq y_{r_{\max}}$.
% \begin{align*}
%      \quad x_{r_{\min}} &\leq x_r  \leq x_{r_{\max}},\\
%      \quad y_{r_{\min}} &\leq y_r \leq y_{r_{\max}}.
% \end{align*}

 Due to the similar structure of $D_{\rm CU}$ and $D_{\rm D2D}$, here too we cannot find a closed-form optimal solution. Accordingly, on similar lines, we evaluate the quantity 
 %\vspace{-2mm}
\begin{equation}    \label{cderi}    
    D_{\rm CU}'(r) = \sum\limits_{n=1}^N \frac{-\lambda_n \beta_{{z_n,UAV}} d(z_n, r)^{-\beta_{{z_n,UAV}}-1}d'(z_n, r)}{(\ln{2})\left(1+{\lambda_n  d(z_n, r)^{-\beta_{{z_n,UAV}}} }\right)} 
\end{equation}
to propose Algorithm \ref{algo2} for finding the best possible solution numerically.  The worst case time complexity of this algorithm is $O(NN_{\text{max}})$.
%It is interesting to note that, while Algorithm \ref{algo1} depends on the location of both the users of every D2D pair, Algorithm \ref{algo2} depends solely on the location of all the CUs.

\begin{algorithm} [!h]
\caption{Algorithm to Solve $\rm P2$}  \label{algo2}
\begin{algorithmic}[1]
    \State \textbf{Initialize} with initial guess $r_0$ such that $r_{\min} < r_0 < r_{\max}$ and $r_0 \neq z_n$ for all $n$, learning rate $\alpha$, tolerance $\epsilon$, and maximum iterations $N_{\text{max}}$.
    \State Set iteration count $k = 0$.
    \Repeat
        \State Compute $D_{\rm CU}'(r)$ by using \eqref{cderi}.
        \State $r_{k+1} = r_k + \alpha D_{\rm CU}'(r_k)$
        %\State Apply constraints:
        \ForAll{$n$}
            \If{$|r_{k+1} - z_n| < \epsilon$}
                \State $r_{k+1} = r_{k+1} + \operatorname{sign}(r_{k+1} - z_n) \cdot \delta$
            \EndIf
        \EndFor
        \State Ensure bounds: $r_{k+1} = \min(\max(r_{k+1}, r_{\min}), r_{\max})$
    \Until{$|D_{\rm CU}'(r_{k+1})| < \epsilon$ or $|r_{k+1} - r_k| < \epsilon$ or $k \geq N_{\text{max}}$}
    \State \Return $r^* = r_{k+1}$ and $D_{\rm CU}(r^*)$
\end{algorithmic}
\end{algorithm}

\subsection{Best Combined Position for RIS and UAV}
\noindent By solving $\rm P1$, Algorithm \ref{algo1} determines the optimal position for the RIS to serve the current set of requesting D2D pairs. Similarly, Algorithm 2 provides solution to $\rm P2$, to determine the best possible position for the UAV to serve as many CUs as possible. However, in general, these optimal positions obtained from Algorithm 1 and Algorithm 2, respectively, are not the same. If we move away from these individually optimal positions, it is expected that the data rate for both D2D users and CUs will decrease. However, since the RIS is mounted on the UAV, we are interested in finding its joint optimal or `near-optimal' position such that the reduction in data rate remains minimal. This is achieved by using the following approach.

% Algorithm 1 results in the optimal positioning of the RIS for maximization of $D_{\rm D2D}$. Consequently, 

Let the optimal position of RIS obtained by Algorithm \ref{algo1} is $r_{\rm D}$. The corresponding average throughput per D2D pair is defined as
%\vspace{-2mm}
\begin{equation}
    D_{\rm D2D}^{\rm avg}(r_{\rm D})=\frac{D_{\rm D2D}(r_{\rm D})}{M}.
\end{equation}
Similarly, we also look at the throughput of the CUs. If Algorithm \ref{algo2} results in the optimal UAV placement $r_{\rm C}$ for the purpose of $D_{\rm CU}$ maximization, the corresponding average throughput per CU is defined as
%\vspace{-2mm}
\begin{equation}
    D_{\rm CU}^{\rm avg}(r_{\rm C})= \frac{D_{\rm CU}(r_{\rm C})}{N}.
\end{equation}
Next, we calculate the ratio of these two quantities as
%\vspace{-2mm}
\begin{equation}
    \phi=\frac{D_{\rm CU}^{\rm avg}(r_{\rm C})}{D_{\rm D2D}^{\rm avg}(r_{\rm D})}. 
\end{equation}
Our objective is to determine the position $ s = (x,y,H)$ such that $T(s)$ is minimized, where
%\vspace{-2mm}
\begin{equation}    \label{d2dcuopt}
   T(s) = \Big|\frac{D_{\rm CU}^{\rm avg}(s)}{D_{\rm D2D}^{\rm avg}(s)} - \phi \Big|.
\end{equation}
Here $D_{\rm CU}^{\rm avg}(s)$ and $D_{\rm D2D}^{\rm avg}(s)$ represents the average throughput per CU and per D2D user pair, respectively, at $s=(x,y,H)$. Moreover, similar to $r_{\min}$ and $r_{\max}$, we also define the limits of $s$ as $s_{\min} = (x_{s_{\min}}, y_{s_{\min}},H)$ and $s_{\max} = (x_{s_{\max}}, y_{s_{\max}},H)$, respectively. Furthermore, we denote the initial value of $s$ as $s_0$, the mid point of the line joining $r_{\rm D}$ (obtained from Algorithm \ref{algo1}) and $r_{\rm C}$ (obtained from Algorithm \ref{algo2}). In order to find the appropriate $s$, we propose Algorithm \ref{algo3}, which is largely based on the well-known coordinate search method \cite{csrch}. The proposed Algorithm \ref{algo3} works by dividing the $xy$ plane into 360 directions (num\_directions), corresponding to the angles from $0^\circ$ to $360^\circ$. Starting with an initial point $s_0$, the algorithm searches along each direction by adjusting the coordinates in that direction. For each angle, it performs a one-dimensional search, by evaluating the objective function at each step (step\_size). The best solution found for each direction is stored, and the algorithm moves in the direction that minimizes the objective function. Finally the algorithm returns the best position $s$, which minimizes the objective function. The time complexity of the algorithm is $O(\rm num\_ directions * \rm max\_steps)$.

\begin{algorithm}
\caption{Algorithm to find the joint best location}  \label{algo3}
\label{alg:directional_search}
\begin{algorithmic}
\State \textbf{Initialize} with initial guess $s_0 = (x_0, y_0,H)$ such that $s_{\min} < s_0 < s_{\max}$, ${\rm num\_directions}$, $\rm step\_size$ and maximum number of steps $\rm max\_steps$.
\State ${\rm best\_solution} \gets (x_0, y_0)$
\State ${\rm best\_value} \gets T(s_0)$

\For{$i = 0$ to ${\rm num\_directions} - 1$}
    \State $\theta \gets \frac{360}{\rm num\_directions} \times i$ \Comment{Calculate direction angle}
    \State $\rm direction \gets (\cos\theta, \sin\theta)$ \Comment{Convert angle to $(dx, dy)$}
    
    \For{$\rm num\_steps = 1$ to $\rm max\_steps$}
        \State $x_{new} \gets x_0 + {\rm direction}[0] \times {\rm step\_size}$
        \State $y_{new} \gets y_0 + {\rm direction}[1] \times {\rm step\_size}$
        \State ${\rm value} \gets T(x_{new}, y_{new})$
        
        \If{$\rm value < best\_value$}
            \State $\rm best\_value \gets value$
            \State ${\rm best\_solution} \gets (x_{new}, y_{new})$
        \EndIf   
    \EndFor  
\EndFor

\State \Return $({\rm best\_solution})$
\end{algorithmic}
\end{algorithm}

\section{NUMERICAL RESULTS}\label{simulation}
\noindent Here, we evaluate the performance of the proposed algorithm and also compare the results with the existing benchmarks \cite{d2dref,curef}.  Authors of \cite{d2dref} and \cite{curef} have individually considered the D2D users and the CUs and subsequently found the optimal positions of the RISs and UAVs, respectively. The simulation parameters considered are as stated in Table \ref{tab:param}.
% In this section, we analyze the performance of the proposed algorithm under various scenarios. To ensure a comprehensive evaluation, we consider different parameter settings, which are listed in Table I. The results are examined to assess the effectiveness of the algorithm across different conditions.
\vspace{-2mm}
\begin{table}[h]
    \centering
    \caption{Simulation Parameters}
    \vspace{-2mm}
    \label{tab:param}
    \resizebox{0.8\columnwidth}{!}{%
    \begin{tabular}{|l|c|}
        \hline
        \textbf{Parameter} & \textbf{Value} \\
        \hline
        Communication region & $300 \times 300$ ${\rm m}^2$ \\
        Number of D2D pairs users & $M = 80$ \\
        Number of CUs & $N = 30$ \\
        Number of obstacles & $45$ \\
        Number of reflecting elements & $R = 250 $ \cite{sau2025priority} \\
        UAV height & $H=25$ m \\
        Transmit power of $x_m$ & $P_{x_m} = 30$ dBm  \cite{akdeniz2014millimeter} \\
        Gain of the transmit antenna & $G_{x_m} = 24.5$ dBi  \cite{akdeniz2014millimeter} \\
        Gain of the receive antenna & $G_{y_m} = 24.5$ dBi  \cite{akdeniz2014millimeter} \\
         Noise power & $N_0 = -100$ dBm \\
        Carrier frequency & $28$ GHz \\
        PL parameter for LoS link & $\alpha_L = 61.2, \beta_L = 2$ \cite{li2022geometric} \\
        PL parameter for NLoS link & $\alpha_N = 72.0, \beta_N = 2.92$ \\
        \hline
    \end{tabular}
    }
    \vspace{-2mm}
\end{table}

% \begin{figure*}[t]
%  \begin{subfigure}[b]{.33\textwidth}
%     \centering
%     \includegraphics[width=0.98\linewidth]{res2.eps}
%     \vspace{-2mm}
%     \caption{}
%     %\caption{No. of groups vs achievable data rate.}
%     \vspace{-2mm}
%     \label{unserved}
% \end{subfigure}
% \begin{subfigure}[b]{.33\textwidth}
%     \centering
%     \includegraphics[width=0.98\linewidth]{res1.eps}
%     \vspace{-2mm}
%     \caption{}
%     %\caption{Patch spacing vs achievable data rate.}
%     \vspace{-2mm}
%     \label{delay}
% \end{subfigure}
% \begin{subfigure}[b]{.33\textwidth}
%     \centering
%     \includegraphics[width=0.98\linewidth]{res3.eps}
%     \vspace{-2mm}
%      \caption{}
%     %\caption{No. of patches in a group vs achievable data rate.}
%     \vspace{-2mm}
%     \label{ee}
% \end{subfigure}
% \caption{\footnotesize  Impact on (a) $\%$ of unserved users, (b) standard deviation of the user delay, and (c) system energy efficiency.}
% \vspace{-6mm}
% \end{figure*}

\begin{figure}[!h]
\centering\includegraphics[width=0.76\linewidth]{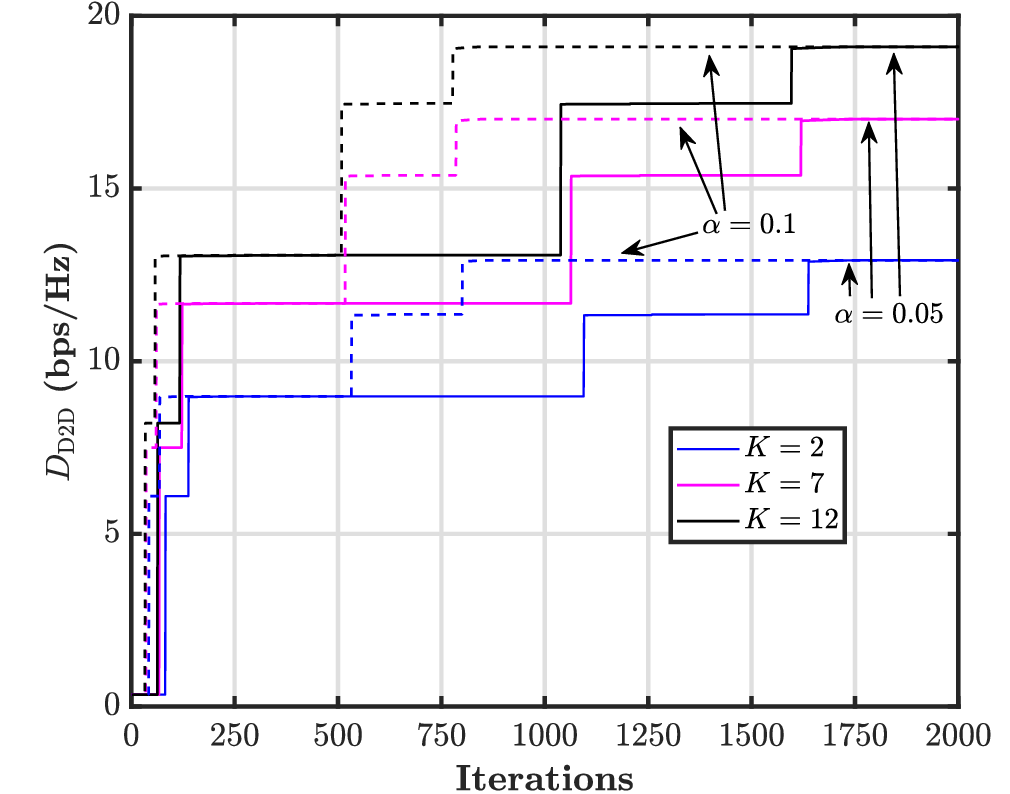}
\vspace{-2mm}
\caption{Convergence of Algorithm \ref{algo1}.}
\label{fig:res1}
\vspace{-4mm}
\end{figure}

Fig. \ref{fig:res1} demonstrates the convergence of the proposed Algorithm \ref{algo1}, for various values of the Rician parameter $K$. Specifically, this algorithm provides the optimal location of the RIS in order to maximize the total throughput of the D2D users. Note that the sole objective of this algorithm is to maximize $D_{\rm D2D}$ without taking care of the CUs present in the environment. This figure illustrates that the proposed algorithm converges in finite time, with the time taken for this purpose being inversely proportional to the learning rate $\alpha$. Moreover, we also observe that the $D_{\rm D2D}$ achieved is higher for higher values of $K$, implying the importance of having a LoS between D2D users and the RIS.
\vspace{-2mm}

\begin{figure}[!h]
\centering\includegraphics[width=0.76\linewidth]{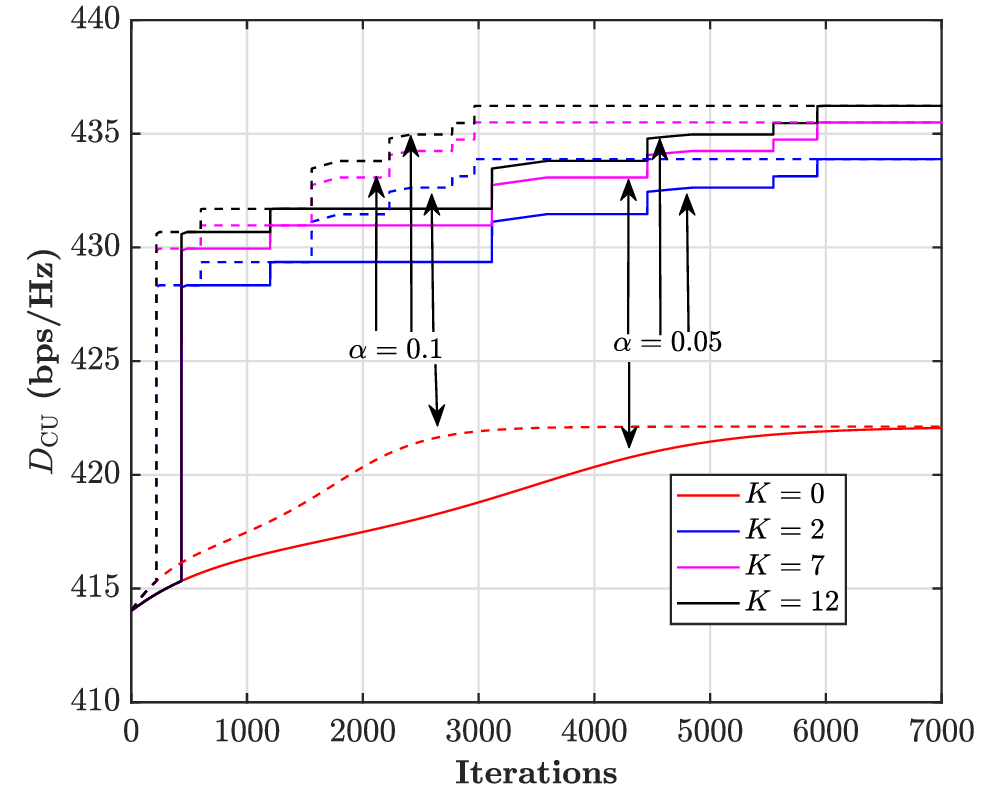}
\vspace{-2mm}
\caption{Convergence of Algorithm $2$.}
\label{fig:res2}
\vspace{-4mm}
\end{figure}

Similarly, Fig. \ref{fig:res2} depicts the convergence of Algorithm \ref{algo2}, which provides the optimal location of the UAV to maximize the total throughput $D_{\rm CU}$ of the active CUs. Even while the earlier claim regarding the contribution of $\alpha$ to the convergence of Algorithm \ref{algo1} and the influence of $K$ is still true, we can also see that the system performance is much worse when $K=0$. This is quite intuitive, as $K=0$ results in a Rayleigh fading scenario without a LoS link between the CUs and the UAV.
However, our objective is always to place the UAV in such a location, which will result in a LoS link for the maximum possible number of active CUs, if not all. 
\vspace{-2mm}

\begin{figure}[!h]
\centering\includegraphics[width=0.76\linewidth]{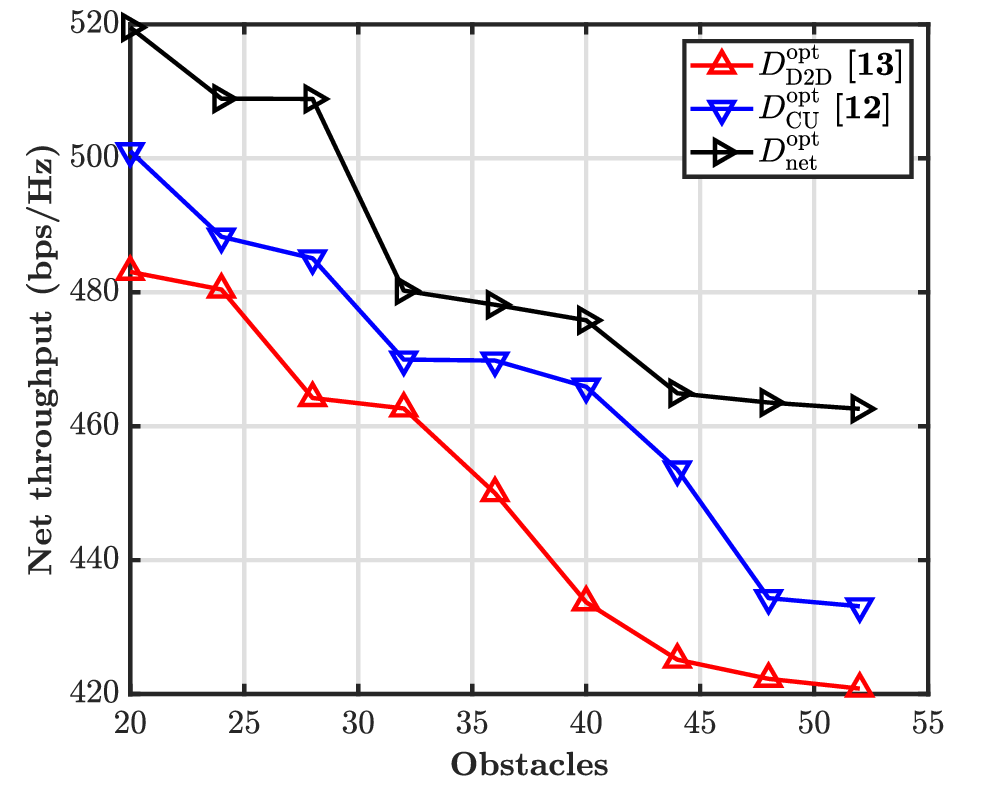}
\vspace{-2mm}
\caption{Impact of obstacles on the system performance.}
\label{fig:res3}
\vspace{-4mm}
\end{figure}

It is important to note that, while Algorithm \ref{algo1} and Algorithm \ref{algo2} may provide different optimal RIS and UAV locations, it is practically not feasible. This is because the RIS is placed on top of the UAV. Therefore, we search for a position, which enhances the net throughput of both the D2D users as well as the CUs at the same time. Algorithm \ref{algo3} finds this position. In Fig. \ref{fig:res3}, we investigate the impact of the obstacles present in the environment on the net throughput $D_{\rm net}$ obtained by \eqref{dnet}. In this figure, $D_{\rm net}^{\rm opt}$ represents the net throughput computed based on the optimal location obtained by Algorithm \ref{algo3}. Here, $D_{\rm D2D}^{\rm opt}$ and $D_{\rm CU}^{\rm opt}$ represent the net throughput obtained by \cite{d2dref} and \cite{curef}, respectively. According to \cite{d2dref}, we first determine the best location for D2D users and compute $D_{\rm D2D}$. We then compute $D_{\rm CU}$ for the CUs using the same location obtained for the D2D users. Finally, we compute $D_{\rm net}$ by using \eqref{dnet} and denote it by $D_{\rm D2D}^{\rm opt}$. In the same way, we compute $D_{\rm CU}^{\rm opt}$ based on the optimal location for CUs as in \cite{curef}. We observe from the figure that irrespective of the framework, the net throughput decreases with increasing obstacles, which is intuitive. However, it is interesting to see that, $D_{\rm net}^{\rm opt}$ is always greater than the other schemes, which aim to individually provide greater throughput to the D2D users or the CUs. This brings out the novelty of our proposed framework, as to individually maximizing $D_{\rm D2D}$ or $D_{\rm CU}$ does not necessarily imply enhancing the total system throughput. Algorithm \ref{algo3} outperforms the others, as it considers the best location for both the D2D users and the CUs. 
\vspace{-2mm}
\vspace{-2mm}
\begin{figure}[!h]
\centering\includegraphics[width=0.76\linewidth]{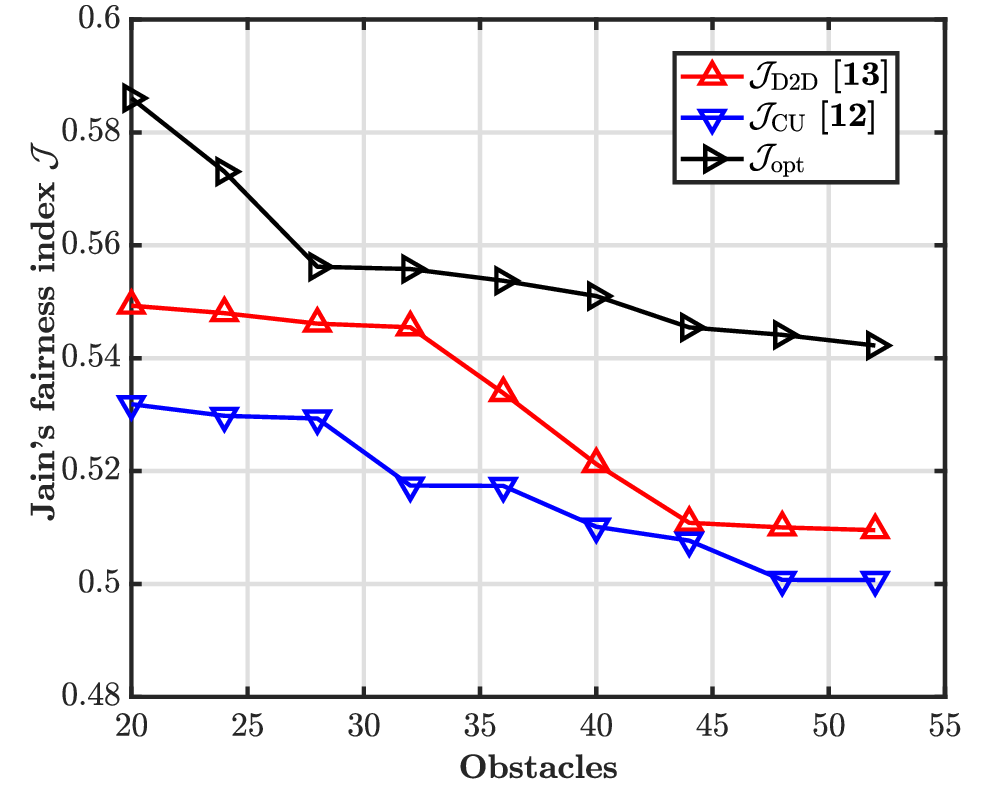}
\vspace{-2mm}
\caption{Impact of obstacles on the system fairness.}
\label{fig:res4}
\vspace{-4mm}
\end{figure}

Finally, we look at the impact of the obstacles on the Jain's fairness  index \cite{jain} $\mathcal{J}_{\rm D2D}$,  $\mathcal{J}_{\rm CU}$ and $\mathcal{J}_{\rm opt}$ computed for the approaches in \cite{d2dref}, \cite{curef} and our proposed approach (Algorithm \ref{algo3}), respectively, in Fig. \ref{fig:res4}. In this figure, we observe that Algorithm \ref{algo3} results in the best performance among the existing benchmarks \cite{d2dref}, \cite{curef}. In other words, our approach is fair to both the D2D pairs and the CUs.

\section{CONCLUSION} \label{conclusion}
\noindent In this work, we investigated the aspect of optimal RIS-mounted UAV placement. Specifically, we consider two types of users, i.e., the D2D pairs and the CUs. We propose algorithms in this direction, which aim to maximize the combined system throughput of both these types of users and not only one of them. The numerical results demonstrate the superiority of the proposed framework in comparison to the existing benchmark schemes in terms of both throughput and fairness.

\bibliography{refs}
\IEEEpeerreviewmaketitle
\end{document}